\documentclass[fleqn,usenatbib]{mnras}

\usepackage{newtxtext,newtxmath}

\usepackage[T1]{fontenc}
\usepackage{ae,aecompl}

\usepackage{graphicx}	
\usepackage{amsmath}	

\usepackage{xcolor}     
\usepackage{hyperref}
\usepackage{subfig}
\usepackage{tablefootnote}
\usepackage{pdflscape}

\usepackage[document]{ragged2e}

\title[Cross-correlation to Constrain IGMF]{Constraints on Large-Scale Magnetic Fields in the Intergalactic Medium Using Cross-Correlation Methods}

\author[Amaral et al.]{
A.D. Amaral,$^{1,2}$\thanks{E-mail: amaral@astro.utoronto.ca}
T. Vernstrom,$^{2,3}$
and B.M. Gaensler$^{2,1}$
\\
$^{1}$David A. Dunlap Department of Astronomy and Astrophysics, University of Toronto, ON, M5S 3H4, Canada\\
$^{2}$Dunlap Institute for Astronomy and Astrophysics, University of Toronto, Toronto, ON, M5S 3H4, Canada\\
$^{3}$CSIRO Astronomy and Space Science, PO Box 1130, Bentley, WA 6102, Australia
}

\date{Accepted XXX. Received YYY; in original form ZZZ}

\pubyear{2015}

\begin{document}
\label{firstpage}
\pagerange{\pageref{firstpage}--\pageref{lastpage}}
\maketitle

\begin{abstract}

Large-scale coherent magnetic fields in the intergalactic medium are presumed to play a key role in the formation and evolution of the cosmic web, and in large scale feedback mechanisms.  However, they are theorized to be extremely weak, in the nano-Gauss regime. To search for a statistical signature of these weak magnetic fields we perform a cross-correlation between the Faraday rotation measures of 1742 radio galaxies at $z > 0.5$ and large-scale structure at $0.1 < z< 0.5$, as traced by 18 million optical and infrared foreground galaxies. No significant correlation signal was detected within the uncertainty limits.   We are able to determine model-dependent  $3 \sigma$ upper limits on the parallel component of the mean magnetic field strength of filaments in the intergalactic medium of $\sim 30 \ \mathrm{nG}$ for coherence scales between $1$ and $2.5 \  \mathrm{Mpc}$, corresponding to a mean upper bound RM enhancement of $\sim 3.8 \ \mathrm{rad/m^{2}}$ due to filaments along all probed sight-lines.  These upper bounds are consistent with upper bounds found previously using other techniques.  Our method can be used to further constrain intergalactic magnetic fields with upcoming future radio polarization surveys.

\end{abstract}

\begin{keywords}
magnetic fields --  intergalactic medium: intergalactic filaments -- large-scale structure of the universe -- methods: statistical -- radio continuum: galaxies
\end{keywords}



\section{Introduction}

Astronomers have been able to detect the presence of magnetic fields across many scales, both Galactic and extra-galactic (see \citealt{Vallee1997,Vallee1998,Vallee2004} for reviews). However, magnetic fields on the largest scales in the Universe remain largely unconstrained \citep{Widrow2002}.  These scales are occupied by filaments, voids, and galaxy clusters, where over-dense filaments connect to form the Universe's large-scale structure known as the cosmic web \citep{Springel2006}.

The dominant gas phase of filaments is in the form of the warm hot ionized medium (WHIM), which contains baryons at high temperatures ($10^5$K $\leq T \leq 10^7$K, \citealt{Ryu2008}) due to shock heating. At low redshifts, the WHIM is thought to contain $\sim 40\%$ of the intergalactic medium (IGM) and the majority of the baryons in the Universe, making it an important component of the Universe to understand \citep{Cen1999,Cen2006,Dolag2006,Dave2010}.
These IGM filaments play a vital role in intergalactic gas feedback processes by providing pristine gases and elements to galaxies to fuel star formation via cold streams \citep{Man2018}. Any large-scale magnetic fields present in filaments would likely have an effect on this process \citep{Klar2012}.  Moreover, most theories that explain how magnetic fields in galaxies form and evolve over cosmic time (such as the $\mathrm{\alpha-\Omega}$ dynamo, see \citealt{Kulsrud2008} for a review) require the presence of an initial weak seed field within the IGM at early times.

It is certain that magnetic fields existed on small-scales in the primordial Universe, due to the presence of currents, though their presence on larger cosmic scales remains theoretical \citep{Grasso2001}. These primordial magnetic fields (PMFs) may have been generated during early phase transitions along phase bubbles  (such as the quantum chromodynamic and electroweak phase transitions, \citealt{Widrow2002}), or shocks present during these transitional phase boundaries could have amplified and generated magnetic fields with coherence scales on the order of the phase bubbles \citep{Kahniashvili2013}.  Big bang nucleosynthesis chemical abundances place bounds on PMFs from $\mathrm{nG}$ to $\mathrm{\mu G}$ \citep{Grasso1995,Cheng1996,Kawasaki2012}. Bounds on the presence of PMFs from the temperature and polarization maps of the cosmic microwave background (CMB) are $ \leq 10 \ \mathrm{nG}$ \citep{Planck2015}.

Cosmic-scale magnetic fields also may have been seeded from astrophysical sources that release large amounts of magnetic flux into the IGM. Examples of such sources include the highly magnetized coherent outflows from active galactic nuclei (AGN), which contain charged particles undergoing acceleration in jets \citep{Furlanetto2001}. The strength of the outflows can stretch magnetic field lines, causing large coherence scales and spreading outwards into the IGM \citep{Furlanetto2001}. In a similar manner, the first starburst galaxies could have generated significant magnetized winds due to rapid star formation, injecting magnetic flux into the surrounding IGM \citep{Kronberg1999}.

The presence of magnetic fields have been inferred observationally for galaxies at high redshifts \citep{Kronberg1982,Wolfe1992,Oren1995,Bernet2008,Mao2017}, indicating that generation mechanisms must be present in the early universe.
Additionally, large-scale $\mathrm{\mu G}$ fields have been detected in the intra-cluster medium (ICM) between galaxies \citep{Kim1991,Feretti1995,Clarke2001,Bonafede2010,Bonafede2013}, and in galaxy cluster haloes on the outskirts of the ICM \citep{Roland1981,Kim1990}.


The difficulty detecting intergalactic magnetic fields (IGMFs) is due to the fact that these fields are theorized to be extremely weak, in the $\mathrm{nG}$ regime, and contain magnetic field reversals \citep{Ryu2008,Cho2009,Akahori2010,Vazza2015,Akahori2016}.
Additionally, the density of relativistic particles is low in these environments \citep{Cen2006,Bregman2007}, making detection via synchrotron emission difficult \citep{Brown2011,Vazza2015b}.

Direct detections of IGMFs may only be possible through data from future telescopes (such as the upcoming Square Kilometre Array and its pathfinders, \citealt{Gaensler2009}). Until then, statistical and indirect methods must be used in order to infer their presence.  Measured properties from these techniques, such as the field strength and coherence scale, can differentiate between magneto-genesis models \citep{Donnert2009,Vacca2015}. 

Statistical methods using Faraday rotation data can be used to place upper bounds on line-of-sight magnetic fields permeating the IGM. Magneto-ionic media are birefringent, thus as linearly polarized light pass through magnetic fields it undergoes Faraday rotation, in which the polarization angle of the radiation rotates along the line-of-sight. The degree of observed rotation is:

\begin{equation}\label{RM_angle_def}
\Delta \Phi =RM \lambda^{2},
\end{equation}

where  $\lambda$ is the observed wavelength, and RM is a proportionality constant called the rotation measure. This RM can be used to infer information about the line-of-sight component of the magnetic fields causing the Faraday rotation.  The RM is given by:

\begin{equation}\label{RM_def}
RM = 812 \int_{z_s}^{0} (1+z)^{-2} n_{e}(z) B_{\parallel}(z) dz \cdot \frac{dl}{dz} \; \mathrm{rad  \; m^{-2}},
\end{equation}

where $z_s$ is the redshift of the emitting polarized source, $n_{e}$ is the column density of free electrons along the line-of-sight to the source measured in $cm^{-3}$, $B_{\parallel}$ is the line-of-sight component of the magnetic field measured in $\mu G$, and the line-of-sight $dl$ is measured in kpc. 

Because the RM is an integrated effect along the line-of-sight from a distant source towards Earth, it contains the rotation measure induced within the source itself ($\mathrm{RM_{\rm intrinsic}}$), an intervening extra-galactic component ($\mathrm{RM_{\rm intervening}}$), a Galactic component ($\mathrm{RM_{\rm Galactic}}$), and a component due to noise ($\mathrm{RM_{\rm noise}}$):

\begin{equation}\label{RM_components}
   \mathrm{ RM = RM_{intrinsic} + RM_{\rm intervening} + RM_{\rm Galactic} + RM_{noise}}.
\end{equation}

\begin{figure*}%
	\centering%
    \includegraphics[width=\textwidth]{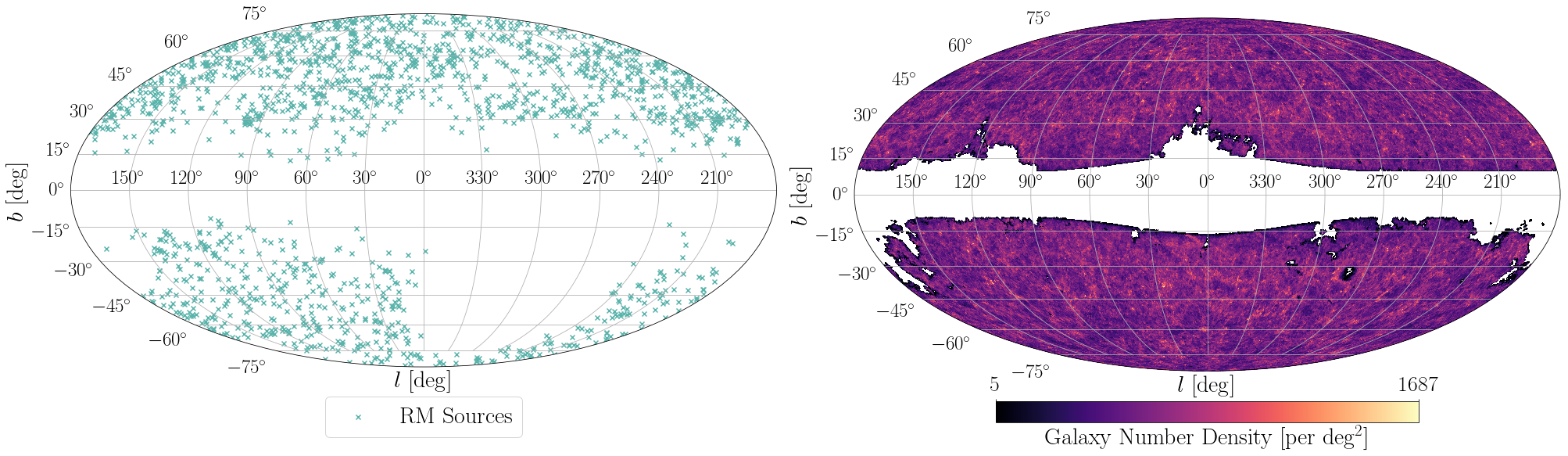}%
    \protect\caption{\textbf{Left}: Locations of 1742 background RM sources from \protect\cite{Hammond2013a} after making selection cuts described in Section \ref{SubSec:Sample_Selection}. \textbf{Right}: 
   Distribution of 18 million foreground galaxies ($0.1<z<0.5$) from WISExSuperCOSMOS.}
    \label{fig:cat_sky_pos}%
\end{figure*}

Various statistical methods have been applied to place upper bounds on large-scale IGMFs, such as quantifying RM growth over redshift for large samples \citep{Blasi1999,Kronberg2008,Bernet2013,Hammond2013a,Xu2014,Pshirkov2016,Neronov2013}, and quantifying RM differences between physical pairs vs. sources that are close on the sky but occupy different redshifts \citep{Vernstrom2019}. In cases where redshift information is unavailable, studies looked at non-physical pairs assuming that z would not be identical \citep{OSullivan2020,Stuardi2020}. 
These techniques have placed upper bounds on IGMFs ranging from $\sim \mathrm{nG}$ to $\sim \mathrm{\mu G}$, depending on the method and assumptions used.

\subsection{Cross-correlation techniques}

Cross-correlation techniques between extragalactic RM-grids (regions of sky with large numbers of RM measurements) and tracers of large-scale structure can be used as another method to extract the IGMF signal.  Because it is difficult to directly observe large scale structure, wide-field galaxy catalogues and galaxy number density grids can be used to trace large-scale structure and the corresponding IGM.  This has been done with success using various wide field multi-wavelength galaxy catalogues (see \citealt{Jarrett2004}, or  \citealt{Gilli2003} for examples). If large-scale magnetic fields exist within the IGM as traced by galaxy densities, any polarized sources behind this IGMF will have an intervening RM contribution due to this magnetic field, as described in equation \ref{RM_components}. If the background RMs are correlated with the foreground galaxy densities on large scales, this suggests the presence of a large-scale magnetic field within the IGM.

A cross-correlation technique was used on mock foreground-subtracted extragalactic RM catalogues 
by \cite{Kolatt1998} to detect features of a PMF power spectrum. \cite{Xu2006} cross correlated RMs in galaxy superclusters (such as Hercules, Virgo, and Perseus), with galaxy densities of the clusters to search for a signal from the IGMF within these structures; they found an upper limit of $\sim 0.3 \ \mathrm{\mu G}$ for coherence scales of 500 kpc.  

More recently, \cite{Lee2009a} used a cross-correlation technique over large regions of the sky ($120^{\circ} < \alpha < 240^{\circ}$,  $0^{\circ} < \delta < 60^{\circ}$) between 7244 extragalactic RMs from \cite{Taylor2009} catalogue, and the sixth data release of the Sloan Digital Sky Survey (SDSS) with photometric redshifts \citep{Oyaizu2008} to trace large-scale structure.  They found a positive correlation between the two quantities at large-scales, indicating a field strength of  $\sim 30\ \mathrm{nG}$.
However, \cite{Lee2009b} withdrew this result because the cross-correlation was found to be spurious. This points to the prospect of detecting such signals, but also demonstrates that the interpretation can be susceptible to statistical errors. This claimed result and its subsequent retraction motivates our study: we here revisit this technique, but with a more careful and thorough approach to systematics and uncertainties.

Some work has been done on what such cross-correlations would look like for large data-sets of simulated RMs of various large scale IGMF strengths, and using galaxy catalogues as tracers for structures on large-scales (such as by \citealt{Stasyszyn2010} and  \citealt{Akahori2014a}).  Given sufficient sample sizes, one can discern between magneto-genesis models given the shape and amplitude of the cross-correlation.

In this work, we place upper bounds on the IGMF by using cross-correlation techniques between the largest RM-redshift catalogue currently available, and an all-sky galaxy catalogue with photometric redshifts. The paper is organized as follows; in Section 2 we describe the methods used to constrain the IGMF signal; in Section 3 we discuss the data used in our analysis; in Section 4 we present the cross-correlation and its results; and in Section 5 we place corresponding upper limits on the magnetic field of the IGM, and discuss how future work with upcoming surveys can improve constraints on large-scale magnetic fields in the IGM.  Throughout the paper we assume a $\Lambda CDM$ cosmology of $H_0 = 67.8\;\mathrm{km\;s^{-1}\;Mpc^{-1}}$ and $\Omega_{M} = 0.308$ \citep{Plank2016_H0}.

\begin{figure*}\centering
\subfloat[]{\label{fig:gal_z_distribution}\includegraphics[width=.5\linewidth]{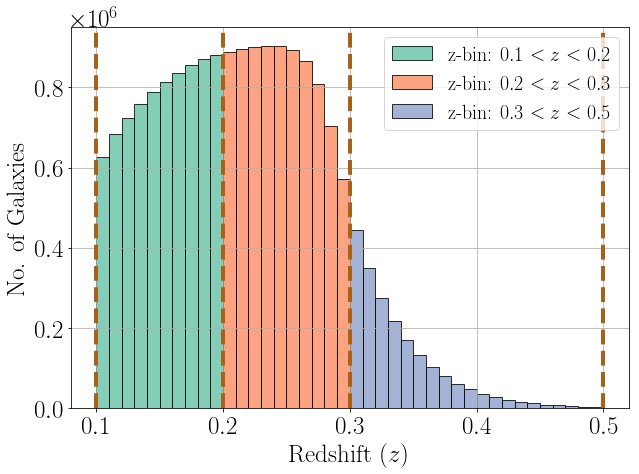}}\par

\subfloat[]{\label{fig:RM_distribution_RM}\includegraphics[width=.5\linewidth]{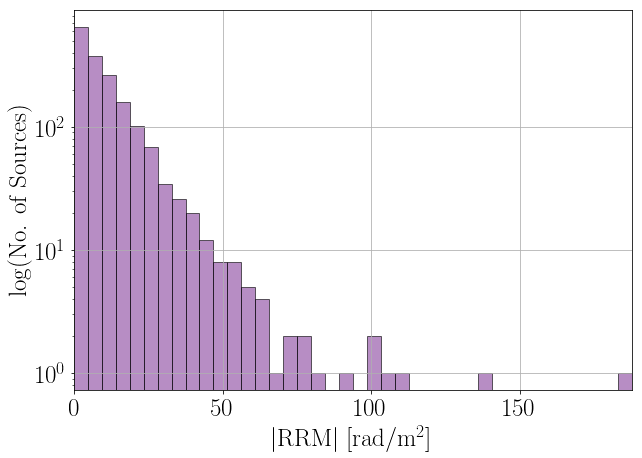}}\hfill
\subfloat[]{\label{fig:RM_distribution_z}\includegraphics[width=.5\linewidth]{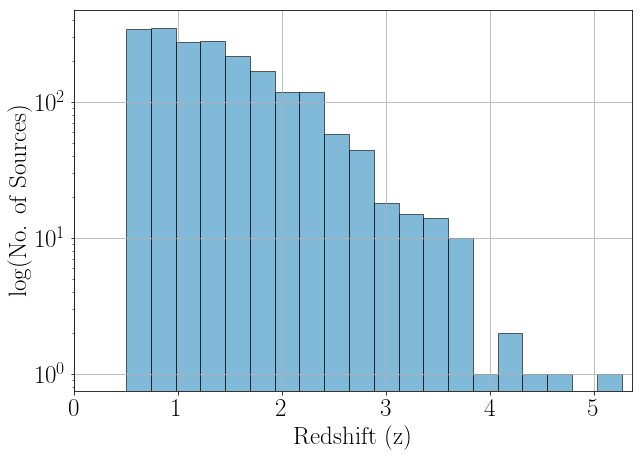}} 
\caption{The distributions of the polarized background RM sources, and the foreground galaxies used in the cross-correlation: \textbf{(a)} The redshift distribution of the galaxies used in this study from WISExSuperCOSMOS. \textbf{(b)} The $\mathrm{|RRM|}$ distribution of the RM sources used, and \textbf{(c)} The redshift distribution of the RM sources used.}
\label{fig:data_distributions}
\end{figure*}

\section{Data}\label{Sec:Data}

\subsection{Rotation Measure Catalogue}\label{SubSec:RM_cat}

We use the \cite{Taylor2009} RM catalogue, which derived RMs from the NRAO VLA Sky Survey (NVSS, \citealp{Condon1998}). The NVSS contains images of Stokes I, Q, U, at 1.4 GHz, at 45 arcsecond resolution for 1.8 million sources at declinations greater than $\delta > - 40 ^{\circ}$ . \cite{Taylor2009} then reprocessed the original NVSS data to determine RMs for 37 543 radio sources with signal-to-noise ratios greater than $8 \sigma$, using two frequency bands centred at 1364.9 MHz and 1435.1 MHz.  Given that only two narrowly spaced frequency bands were used to determine the RMs, the values derived might be prone to errors (such as the $n\pi$ ambiguity; see Section \ref{SubSec:Uncertainties} for further discussion).  This is currently still the largest single catalogue of rotation measures to date.

For our work, the RM sources must be at large redshifts such that their radiation passes through the foreground galaxies that trace large-scale structure. 
 To determine if a source is a background source, the redshift is needed - this information is not available in \cite{Taylor2009}.  \cite{Hammond2013a} cross-matched the \cite{Taylor2009} catalog with optical surveys and online databases (such as SDSS DR8; \citealt{Aihara2011}), and databases, such as SIMBAD \citep{Wenger2000}\footnote{http://simbad.u-strasbg.fr/simbad/} and NED\footnote{The NASA/IPAC Extragalactic Database (NED) is funded by the National Aeronautics and Space Administration and operated by the California Institute of Technology. More information can be found at: https://ned.ipac.caltech.edu}, to obtain spectroscopic redshifts. This yields a sample of 4003 sources with RMs and spectroscopic redshifts, which allows us to ensure that the RM sources are indeed background sources, or at higher redshifts, than the galaxies (see Section \ref{SubSec:Data_GalCat}) for cross-correlating. The left panel in Figure \ref{fig:cat_sky_pos} shows the sky distribution of RMs used for this study.

\subsection{Galaxy Catalogue}\label{SubSec:Data_GalCat}

We use the WISExSuperCOSMOS \citep{Bilicki2016} survey as our tracer for foreground large-scale structure.  This catalogue is the product of cross-matching between the mid-IR Wide-field Infrared Survey Explorer (WISE, \citealp{Wright2010}) and the optical SuperCOSMOS \citep{Hambly} all-sky surveys, and contains 20 million cross-matched galaxies after filtering for quasars and stars and other tests. 
To generate accurate photometric redshifts, \cite{Bilicki2016} cross-matched their WISExSuperCOSMOS catalogue with the Galaxy and Mass Assembly-II (GAMA-II) survey \citep{Driver2009}, an extragalactic (clean from stars and quasars) spectroscopic survey that includes 193,500 overlapping sources with WISExSuperCOSMOS.  They used this as a training set for the ANNz package, an artificial neural network package that assigns photometric redshifts to a sample given a training set with both photometric and spectroscopic redshifts. \cite{Bilicki2016} were thus able to determine photometric redshifts for all 20 million sources. 

We require redshifts for the galaxies to ensure that they are indeed foreground to the RM sources discussed in Section \ref{SubSec:RM_cat}. The WISExSuperCOSMOS catalogue was used due to the fact that it provides redshifts necessary for this study (with a median galaxy redshift $z \approx 0.2$). Compared to other galaxy surveys, WISExSuperCOSMOS also has many sources and covers at least $70\%$ of the sky, allowing for a good tracer of large-scale structure. The right panel of Fig. \ref{fig:cat_sky_pos} shows the galaxy density distribution used for this study.

\begin{figure*}
	\centering
    \includegraphics[width=0.8\textwidth]{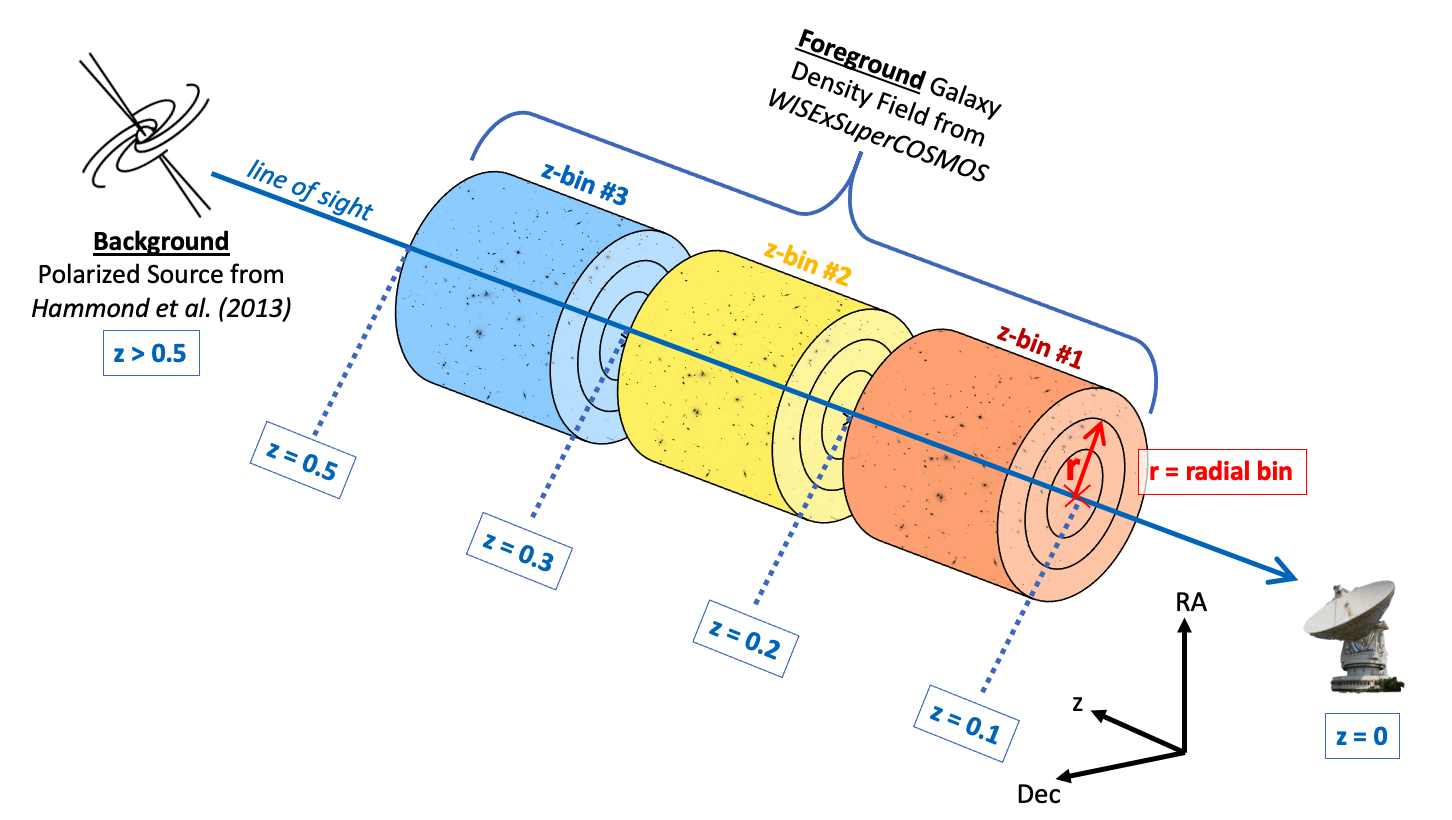}
    \caption{An illustration of our cross-correlation technique. An extra-galactic polarized background source ($z>0.5$) whose radiation passes through or near foreground galaxies ($0.1 < z < 0.5$, separated into three redshift bins) is observed by a telescope on Earth. In order to probe correlations on large-scales, we use 8 linearly spaced radial bins shaped as annuli proceeding outward from the line-of-sight of each RM source, out to an impact parameter of 2.5 Mpc.  Only 3 of the 8 radial bins, $r$, are shown. All images used in this illustration are covered under creative commons: \href{http://thenounproject.com/term/quasar/1667381/}{"Quasar"} by \href{http://thenounproject.com/Vntole/}{Anthony Ledoux} is licensed under \href{http://creativecommons.org/licenses/by/3.0/us/legalcode}{CC BY 3.0}, \href{https://commons.wikimedia.org/wiki/File:70-\%D0\%BC_\%D0\%B0\%D0\%BD\%D1\%82\%D0\%B5\%D0\%BD\%D0\%BD\%D0\%B0_\%D0\%9F-2500_(\%D0\%A0\%D0\%A2-70).jpg}{"70-m aerial P-2500 (RT-70 radio telescope)"} by S. Korotkiy is licensed under \href{http://creativecommons.org/share-your-work/licensing-considerations/compatible-licenses}{CC BY-SA}, and \href{http://www.spacetelescope.org/images/potw1819a}{"HST image of galaxy cluster RXC J0032.1+1808"} by \href{http://commons.wikimedia.org/wiki/European_Space_Agency}{ESA/Hubble \& NASA, RELICS} is licensed under \href{http://creativecommons.org/licenses/by/4.0}{CC BY 4.0}.}
    \label{fig:CCF_situation}
\end{figure*}

\section{Cross-Correlation Analysis}

\subsection{Sample Selection}\label{SubSec:Sample_Selection}
To cross-correlate between RMs and large-scale structure, we must ensure that the polarized radio sources are background sources whose radiation pass through the IGM of the large-scale structure traced by foreground galaxies.  
The redshift distribution of the WISE galaxies severely drops off for $z>0.5$ as shown in Figure \ref{fig:gal_z_distribution}. Thus we set $z=0.5$ as the redshift limit; defining RM sources with $z< 0.5$ as foreground and $z\geq 0.5$ as background sources. 
\cite{Bilicki2016} found that the photometric redshifts of the galaxy catalogue for galaxies with $z<0.1$ contained higher fractional errors and contamination of foreground stars. We thus require galaxies to have redshifts between $0.1 < z < 0.5$, leaving $18,189,238$ foreground galaxies.

Due to uncertainties associated with photometric redshifts of the foreground galaxies, we choose to use redshift bins of widths of $\Delta z = 0.1$ for the first two redshift bins (as recommended by \citealt{Bilicki2016}) and $\Delta z=0.2$ for the final redshift bin for our analysis.
To ensure a large enough number of sources in each bin, we chose to bin the galaxies in the following manner: $0.1 < z < 0.2$ (containing $7,840,929$ galaxies), $0.2 < z < 0.3$ (containing $8,336,519$), and $0.3 < z < 0.5$ (containing $2,011,790$ galaxies). There is a total of $18,189,238$ galaxies across all three redshift bins.
The mean redshift of the galaxies in each bin is: $z_{\mathrm{mean}} = 0.15, 0.247,$ and $0.34$, respectively.

We use a \texttt{HEALPix}\footnote{http://healpix.sourceforge.net} \citep{Gorski2005} gridding scheme and \texttt{healpy}\footnote{http://github.com/healpy/healpy/} \citep{Zonca2019} to split-up the all-sky foreground galaxies into a number density grid using spherical projection pixels of equal surface area. We use $nside = 128$, which corresponds to each pixel occupying $\mathrm{0.21 \  deg^2}$ on the sky. The resulting all-sky number density grid of the remaining foreground galaxies can be seen in Figure \ref{fig:cat_sky_pos}.

 Blank cells in the right panel of Figure \ref{fig:cat_sky_pos} are due to the presence of the Galactic foreground in the survey coverage, or due to empty grid cells that contain no galaxies.  We then removed any RM sources that fell into empty grid cells, and any RM source within 2.5 co-moving Mpc of an empty cell or edge of the WISE galaxy number density map - leaving us with 2229 background RM sources.

\cite{Akahori2016} found that magnetic fields due to galaxy clusters tend to dominate the RM along the line-of-sight, which hence drown out any signal due to weaker magnetic fields, such as those in filaments.  Therefore, RMs that intersect cluster sight-lines must be removed from such studies.  We used the \cite{PlanckClusterCat} Sunyaev-Zeldovich (SZ) cluster catalogue to check for RM sources that pass within sight-lines of clusters.  This cluster catalogue is the largest all sky-catalogue, and is well sampled at $z<1$.  We found that only one RM source from our sample fell within $2 R_{500}$ within a cluster line-of-sight, where $R_{500}$ is defined as the radius for which the density of the cluster is 500 times the critical density and is typically taken to be a proxy for the radius of a galaxy cluster. We removed this source.

\cite{Ma2019b} found that the \cite{Taylor2009} RMs were not properly calibrated for off-axis polarization leakage. From re-observing 23 sources \cite{Taylor2009} sources, and using simulations to quantify this effect, they found that off-axis leakage mostly affects sources with low fractional polarization ($p$), and recommend discarding sources with $p<1\%$.  We used this recommendation and applied a conservative cut to our sample to dispose of sources with $p<1.5\%$ (334 sources) to account for off-axis leakage effects.  This left 1894 sources. We checked using cuts on fractional polarization for $1\%$, $1.5\%$, and $2\%$, the cross correlation remained the same within a factor of 2.5 to the cross correlation if we omit a fractional polarization cut. Discarding sources with $p<1.5\%$ allowed us to balance cutting out sources with uncertain off-axis leakage, while also preserving a larger sample size to perform adequate statistics.




\subsection{Residual Rotation Measures}\label{SubSec:residual_RM}
As per equation \ref{RM_components}, the rotation measure is a cumulative effect along the line-of-sight, which contains a component due to Galactic rotation measure (GRM) foregrounds.  The GRM can be subtracted to obtain a residual RM (RRM). From equation \ref{RM_components}, the RRM gives a better representation of $\mathrm{RM_{intervening}}$, the value we wish to extract for this study, although it still also contains components due to $\mathrm{RM_{intrinsic}}$ and $\mathrm{RM_{noise}}$. Because $\mathrm{RM_{Gal}}$ is typically characterized by smooth RMs over large regions of sky and tends to dominate the RM component of the source \citep{Simard-Normandin1980,Leahy1987,Schnitzeler2010}, if not removed it could affect large-scale correlations. $\mathrm{RM_{noise}}$ and $\mathrm{RM_{intrinsic}}$ are unique to each RM source so we do not expect these values to affect our correlation, allowing us to isolate $\mathrm{RM_{intervening}}$ in our correlations. \cite{Oppermann2015} generated an all-sky catalogue of Galactic rotation measures using simulations to differentiate the Galactic component from previously known extragalactic RM sources to obtain the GRM for all sight-lines. 
 
In removing the Galactic foreground contribution, we want to avoid using sight-lines where the uncertainty in the estimate of GRM is worse than the uncertainty in the RM from other contributions. We therefore only use sight-lines that meet the requirement:

\begin{equation}\label{foreground_eq}
\left(1.22\sigma_{RM}\right)^{2} + \sigma_{intrinsic}^2 \geq \sigma_{GRM}^2,
\end{equation}

where $\sigma_{GRM}$ is the uncertainty in the GRM as computed by \cite{Oppermann2015}, $\sigma_{RM}$ is the observed RM uncertainty from \cite{Taylor2009}, $\sigma_{intrinsic} = 6.9 \ \mathrm{rad m^{-2}}$ is the standard deviation on $\mathrm{RM_{intrinsic}}$ estimated by \cite{Oppermann2015}, and we correct published values of $\sigma_{RM}$ by a factor of 1.22 as recommended in \S 4.2.1 of \cite{Stil2011}.

 We subtract the GRM derived by \cite{Oppermann2015} from the 1894 \cite{Taylor2009} RMs  to obtain the RRMs for each source only if equation \ref{foreground_eq} is satisfied: 152 sources did not meet this criteria.  We therefore removed these sources for a final sample of 1742 background RM sources.

The sign attributed to RMs is due to the orientation of the magnetic field causing the rotation.  Because we are not concerned with the direction of the field, the values we use for cross-correlation are $|RRM|$. The $|RRM|$ distribution of the 1742 background RM sources can be found in Figure \ref{fig:RM_distribution_RM}, and the corresponding redshift distribution of these sources can be found in Figure \ref{fig:RM_distribution_z}. The sky distributions of the RM catalogue and galaxy number density grid after applying our selection criteria are shown in Figure \ref{fig:cat_sky_pos}.

\begin{figure*}
	\centering
    \includegraphics[width=\textwidth]{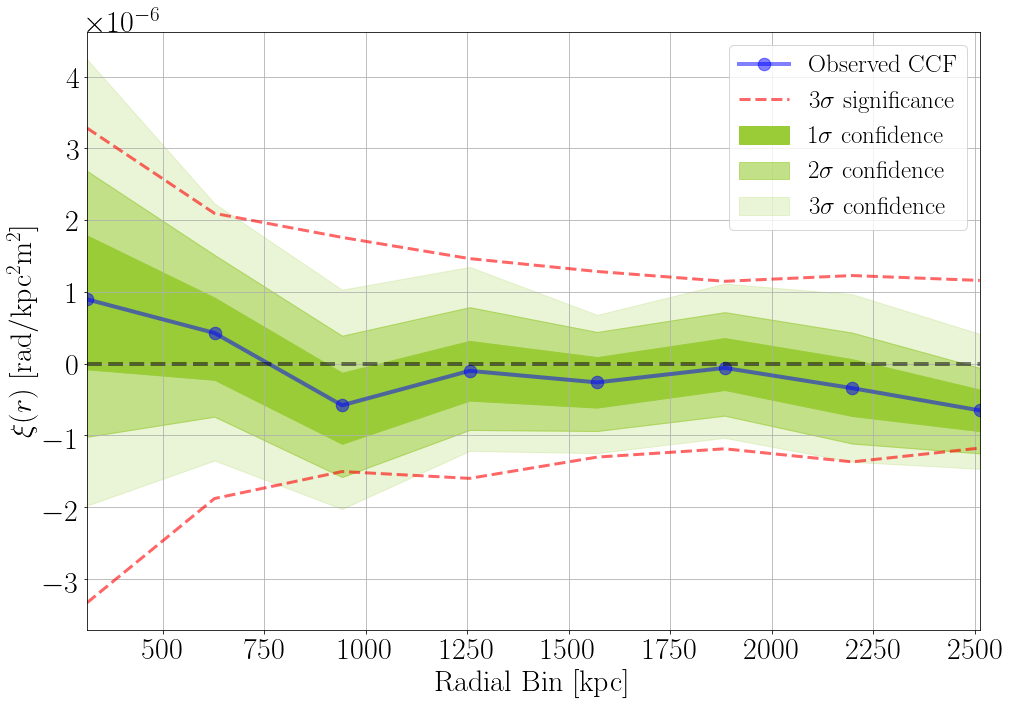}
    \caption{The cross-correlation function (CCF) between the $|RRM|$s of 1742 background sources from \protect\cite{Hammond2013a} and 18 million foreground galaxies selected from \protect\cite{Bilicki2016}. The observed CCF is shown as the solid blue line. The red dashed lines mark the $3\sigma$ significant level of the cross-correlation, while the green coloured regions mark the confidence intervals of the cross-correlation.}
    \label{fig:ccf_final_plot}
\end{figure*}

\subsection{Cross-Correlation Function}\label{SubSec:CC_Function}

A large-scale magnetic field within the filamentary IGM (as traced by the foreground galaxy distribution) will create an enhanced RM along that line-of-sight (extragalactic component, $\mathrm{RM_{intervening}}$, in equation \ref{RM_components}) for each polarized background source.  Probing impact parameters at various distances from the line-of-sight of the RM source allows us to quantify the extent of influence of the field present in the foreground galaxy field.

We wish to probe correlations at large-scales, and thus use 8 linearly spaced radial impact parameter bins extending from the polarized background source between $\sim 500 \ \mathrm{kpc}$ to $2.5 \ \mathrm{Mpc}$ on the sky.  These scales were chosen because the typical thickness of IGM filaments are typically 1-2 Mpc \citep{Ratcliffe1996,Doroshkevich2004,Wilcots2004}, therefore we sample 500 kpc on either side of this range.
Additionally this choice of radial bin scales ensures that the smoothing scale of the GRMs (see section \ref{SubSec:residual_RM}) used will not affect the cross-correlation. The GRM smoothing scale does not affect scales smaller that $\sim 1 \ \mathrm{deg}$ \citep{Oppermann2012}, and angular diameter distance for 1 degree (6.9 Mpc at $z=0.1$) is larger than our largest radial bin of 2.5 Mpc.
The coherence scale of the magnetic field is considered to be the scale within which the magnetic field remains uniform with no field reversals. If a signal is observed at a given scale, $r$, this suggests that the magnetic field is coherent at this scale.

 We then can calculate the cross-correlation function, $\xi (r)$, between the background RRMs and the foreground galaxies using,

\begin{equation}\label{ccf_eq}
\begin{split}
  \xi (r) = \sum_i \sum_j \frac{1}{(1+\overline{z_j})^2} &\left( \frac{N_{gal,i}(r,z_j) - \left< N_{gal}(r,z_j)\right>}{A(r)} \right) \\ & \cdot \left( |RRM|_i - \left<|RRM|\right> \right).
  \end{split}
\end{equation}

The index i represents each individual polarized background source with RRM of $|RRM_i|$, and index $j$ represents the 3 redshift bins of the foreground. $\left<|RRM|\right>$ is the average residual rotation measure of all of the RRM sources, where $\left<|RRM|\right> = 11.1 \ \mathrm{rad/m^{2}}$. Here, $N_{gal,i}(r,z_j)$ is the number of galaxies in redshift bin $z_j$ and at radial impact distance $r$ from the line-of-sight of $RRM_i$, while $\left< N_{gal} (r,z_j)\right>$ is the predicted number of galaxies in each redshift and radial bin around each RRM source. 
The predicted numbers are determined by assuming a Poisson distribution in each redshift bin and calculating the expected number of sources given the on-the-sky area, $A(r)$, of each radial bin, $r$.
As seen in equation \ref{RM_def}, there is a redshift contribution that affects the RM.  The foreground galaxies span a substantial redshift range, thus we un-weight this redshift dependence in equation \ref{ccf_eq} by including the term $\left(1+\overline{z_j}\right)^{-2}$, where $\overline{z_j}$ is the average redshift of that bin.

We are interested in quantifying if any deviations from the average number of galaxies around each RM source correlates with deviations from $\left<|RRM|\right>$.  A visual representation of our cross-correlation method can be found in Figure \ref{fig:CCF_situation}.

\subsection{Results}
\subsubsection{Cross-Correlation of Data}\label{SubSec:CCF_result}
We use equation \ref{ccf_eq} to calculate the cross-correlation function between background RRM sources and foreground galaxies. The resulting value of $\xi(r)$ as a function of r is shown in Figure \ref{fig:ccf_final_plot}.
No strong signal is seen at any value of $r$. There is a slight positive signal in the lowest radial bins, with a decreasing trend with increasing $r$. In order to determine if any significant detection is made, we determined uncertainties as detailed in the following subsection.

\subsubsection{Uncertainties}\label{SubSec:Uncertainties}
In this section, we calculate the confidence and significance levels of the cross-correlation result from section \ref{SubSec:CCF_result}. We define the confidence levels as the intervals in which we are confident that the resultant cross-correlation falls within, whereas the significance levels define how significant the cross-correlation is above a null signal.

To determine the confidence levels on our cross-correlation result, we use a boot-strap method and randomly re-sample the $1742$ RRM sources, with replacement.  We do this for 1000 realizations computing a cross-correlation for each realization, then generating $1\sigma$, $2 \sigma$, and $3 \sigma$ percentiles as the confidence levels on the cross-correlation. 

To obtain the significance levels of the cross-correlation, we produce a null cross-correlation signal.  To do this, we generate random background radio sources randomly located on the celestial sphere such that the correlation between them and the foreground galaxy density should average to zero in all radial bins. 
We mask out any values that fall within the Galactic plane ($-20^{\circ} < b < 20^{\circ}$).

Using the method described in the preceding paragraphs, we calculate random locations on the celestial sphere for 2,674,800 new sight-lines.  From these new locations, we select 1742
and assign each a random RRM from \cite{Hammond2013a} catalogue, using bootstrap resampling to form a set of random locations and RRMs.  We then use these bootstrapped 1742 RRMs to compute the cross-correlation function for 1000 different realizations. 
  We then take the $1 \sigma$, $2 \sigma$, and $3 \sigma$ percentiles from the 1000 random cross-correlation realizations to obtain the significance of the result. 

The final cross-correlation function can be seen in Figure \ref{fig:ccf_final_plot}, where the error bars for various confidence levels are plotted in green, and the $3 \sigma$ significance level of the result (3$\sigma$ confidence level for the null signal) is plotted as red dashed lines. 


\section{Discussion}

The cross-correlation between the foreground galaxies and background RM sources (Figure \ref{fig:ccf_final_plot}) is consistent with zero within $3\sigma$.  Although no significant signal is detected, the cross-correlation function can still be used to obtain upper limits on the magnetic field of the WHIM within the IGM.  We use the significance levels (see Section \ref{SubSec:Uncertainties}) on the cross-correlation function determined from the data to constrain an upper bound on the parallel component on large-scale magnetic fields contained in the WHIM.

\subsection{Predicted RM model}\label{SubSec:Predicted_RM}

To place an upper bound on the magnetic field in the WHIM from our cross-correlation, we first model the rotation measure generated by the WHIM within the foreground galaxy density field, $RM_{fil}$.
We adopt a model for $RM_{fil}$ that is derived from the RM definition (equation \ref{RM_def}):

\begin{equation}\label{eqn:predicted_RM_initial}
RM_{fil, i} = 812 \  \sum_j  \frac{1}{(1+\overline{z_j})^2} n_{e,fil} (z_j) B_{\parallel,fil}(z_j) \frac{1}{\sqrt{N_R(z_j)}} \Delta l_{fil} (z_j),
\end{equation}

where $RM_{fil,i}$ is the observed RM induced by foreground filaments over line-of-sight, $i$. We now explain the components of this model.

The line-of-sight integral in the RM equation is represented by a summation $\Sigma_j$ over each redshift bin $j$. Therefore, the differential path length $dl(z_j)$ is approximated by $\Delta l(z_j) = l(z_{j+1}) - l(z_j)$, where  $l(z_j)$ is the co-moving distance to redshift bin $z_j$:

\begin{equation}\label{eqn:comoving_dist}
l(z_j) = \frac{c}{H_0 (1+ \overline{z_j}) \sqrt{\Omega_M(1+\overline{z_j})^3 + \Omega_{\Lambda} + (1-\Omega_M - \Omega_{\Lambda})(1+ \overline{z_j})^2}}.
\end{equation}
The foreground galaxy densities trace filaments, although filaments do not occupy the entire line-of-sight. To account for this we introduce a line-of-sight volume filling factor for filaments, $f$,  which can range between 0 and 1, and is not a function of z. The line-of-sight path length passing through filaments is then $\Delta l_{fil}(z_j) = f \Delta l(z_j)$.

We include magnetic field reversals in our model, where the magnetic field changes direction over a coherence scale, $L_c$.  We model field reversals as a random walk process \citep{Cho2009}.  This has the effect of reducing the observed RM by $1/\sqrt{N_R}$, where $N_R$ is the number of field reversals over the line-of-sight.  Assuming that $L_c < f \ \Delta l(z_j)$, where $f \ \Delta l(z_j)$ is the total line-of-sight covered by filaments within redshift bin $z_j$, the total number of field reversals along the line-of-sight passing through filaments in redshift bin $z_j$ is then:

\begin{equation}\label{eqn:field_reversals_final}
N_{R} (z_j) = \frac{f \  \Delta l(z_j)}{L_c}.
\end{equation}

The $RM_{fil}$ of the sources will be affected by the number of free electrons contained in filaments along the line-of-sight, $n_{e,fil}(z_j)$.  We assume that the free electron density scales with the relative number density of galaxies:

\begin{equation}\label{eqn:ne_model}
    n_{e,fil}(z_j) = \overline{n_{e,fil}}\left( \frac{n_{gal,2.5,i}(z_j)}{\left< n_{gal,2.5}(z_j)\right> }  \right),
\end{equation}
where $\overline{n_{e,fil}}$ is the average number of free electrons in filaments. 
The weighted galaxy count within an impact parameter of 2.5 Mpc of line-of-sight $i$ and redshift bin $z_j$ is given by $n_{gal,2.5,i}(z_j) = \sum_k N_{gal,i}(r_k,z_j) \times w(r_k)$, where $N_{gal,i}(r_k,z_j)$ is the number of galaxies within radial impact parameter bin $r_k$ and redshift bin $z_j$, and $w(r_k)$ is the radial weighing function evaluated at radial bin $r_k$. $\left< n_{gal,2.5}(z_j)\right>$ is the average weighted galaxy count within an impact parameter of 2.5 Mpc in redshift bin $z_j$ for all $i$ sight-lines.  Therefore, the quotient of these weighted galaxy count quantities represent whether line-of-sight $i$ is in an under- or over-dense region.

We use the radial weighting of a \cite{King1972} profile, given by: $w(r)= \left(1+\frac{r^2}{r_c^2}\right)^{\frac{-3 \beta}{2}}$, where $\beta = 0.7$ \citep{Kronberg2016}, and we choose a scaling length of $r_c = 1 \ \mathrm{Mpc}$ to correspond to the characteristic scale of filament widths \citep{Ratcliffe1996,Doroshkevich2004,Wilcots2004}. 
Including a weight function allows for galaxies closer to the line-of-sight to contribute more to the $RM_{fil}$ along that sight-line than galaxies that are farther away.  This is a reasonable assumption as the magnetic field and $n_e$, and hence the resulting RM, will decrease as $r$ decreases away from regions of high density.

We assume a simple magnetic field case, in which a large-scale coherent magnetic field, $B_{fil}$, is already in place at redshifts $z > 0.5$, and at $0.1 < z < 0.5$ the field is frozen into the plasma.  Thus the magnitude of the magnetic field of the IGM will scale with $n_e$ (electron number densities) present in the foreground galaxy density. We do not consider magnetic energy injected on large-scales by compact astrophysical sources (such as AGN).  This is a reasonable assumption as AGN activity becomes less common in the $z<2$ universe \citep{Madau2014}.  As such, this model is independent of magneto-genesis mechanisms.

We assume that the filaments are randomly distributed and cylindrical in shape (in the $h>>R$ regime, where $h=$ height, and $R=$ radius of a cylindrical filament). 
As in \cite{Blasi1999}, we assume that the magnetic fields and filaments obey both flux conservation and mass conservation, to derive:  

\begin{equation}\label{eqn:Bpara_model}
B_{\parallel, fil}(z_j) = B_{0,\parallel, fil} \left( \frac{n_{gal,2.5,i}(z_j)}{\left< n_{gal,2.5}(z_j)\right> } \right)^{1/2},
\end{equation}

where $B_{0,\parallel, fil}$ is the co-moving mean magnetic field strength in filaments. We assume that for scales $< 2.5 \ \mathrm{Mpc}$, IGM filaments are gravitationally bound structures, and thus equations \ref{eqn:ne_model} and \ref{eqn:Bpara_model} contain no direct cosmological dependence.

Including the above assumptions, equation \ref{eqn:predicted_RM_initial}, the model for the predicted RM due to filaments in a foreground galaxy density grid becomes:

\begin{equation}\label{eqn:RM_model_full}
\begin{split}
    RM_{fil,i} = \ 812 \sum_j  &\frac{f^{1/2}}{(1+\overline{z_j})^2} \overline{n_{e,fil}} \ B_{\parallel,fil,0}  \  \\  & \times \left( \frac{n_{gal,2.5,i}(z_j)}{\left< n_{gal,2.5}(z_j)\right> }\right)^{3/2} \sqrt{\frac{L_c}{\Delta l(z_j)}} \Delta l(z_j).
    \end{split}
\end{equation}

\subsection{Upper Bounds}\label{SubSec:generating_upper_bounds}

\subsubsection{Generating Upper Bounds}

\begin{figure*}
    \centering
    \includegraphics[width=\textwidth]{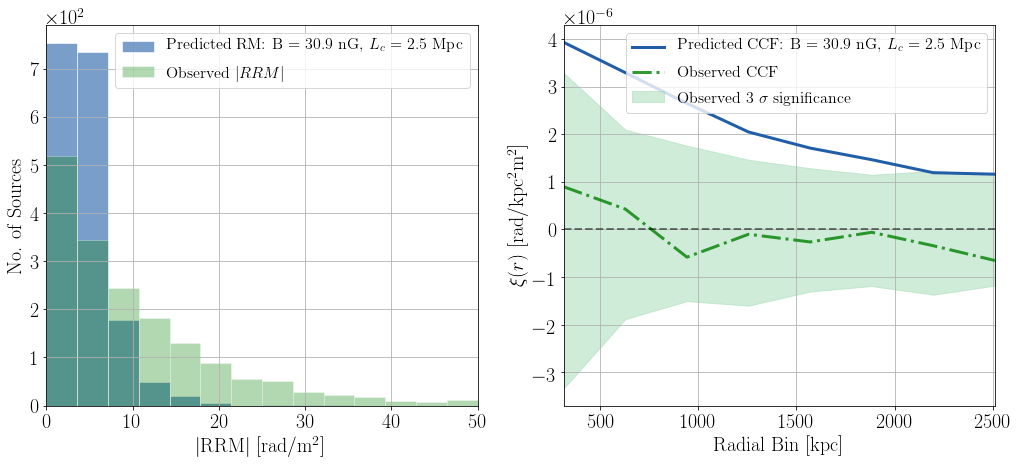}
    \caption{\textbf{Left:} Distribution of predicted RM values compared to observed $|RRM|$ values used, generated using the model given in equation \ref{eqn:RM_model_full}, using $\overline{n_{e,fil}} = 10^{-5} \ \mathrm{cm^{-3}}$, $f=0.06$, and $L_c = 2500 \ \mathrm{kpc}$.  \textbf{Right:}  The resulting cross-correlation function from the observed and predicted RMs shown in the left panel. The green shaded region corresponds to the red dashed lines indicating a $3\sigma$ significance level in Figure \ref{fig:ccf_final_plot}.}
    \label{fig:predicted_RRM_CCF_distribution}
\end{figure*}

\begin{table*}
\begin{tabular}{|l|l|l|l|}
\hline
Coherence Scale ($L_c$) & $f^{1/2} \  \overline{n_{e,fil}} \ B_{\parallel, fil,0}$ Upper Bound &  $B_{\parallel, fil,0}$ Upper Bound & $\left<RM_{fil}\right>$ Upper Bound \\ \hline
$0.5 \ \mathrm{Mpc}$ & $1.1 \times 10^{-7}\ \mathrm{ \mu G \ cm^{-3}}$ & $44 \ \mathrm{nG}$ & $3.0 \pm 1.9 \ \mathrm{rad/m^2}$\\ \hline
$1.0 \ \mathrm{Mpc}$ & $7.9 \times 10^{-8}\ \mathrm{ \mu G \ cm^{-3}}$ & $32 \ \mathrm{nG}$ & $3.1 \pm 2.0 \ \mathrm{rad/m^2}$\\ \hline
$1.5 \ \mathrm{Mpc}$ & $7.3 \times 10^{-8}\ \mathrm{ \mu G \ cm^{-3}}$ & $30 \ \mathrm{nG}$& $3.5 \pm 2.3 \ \mathrm{rad/m^2}$ \\ \hline
$2.0 \ \mathrm{Mpc}$ & $6.6 \times 10^{-8}\ \mathrm{ \mu G \ cm^{-3}}$ & $27 \ \mathrm{nG}$& $3.7 \pm 2.4 \ \mathrm{rad/m^2}$\\ \hline
$2.5 \ \mathrm{Mpc}$ & $7.6 \times 10^{-8}\ \mathrm{ \mu G \ cm^{-3}}$ & $31 \ \mathrm{nG}$& $4.7 \pm 3.0 \ \mathrm{rad/m^2}$ \\ \hline
\end{tabular}
\caption{Table containing the results for constraints on the magnetic field within filaments, using the predicted RM model (equation \ref{eqn:RM_model_full}) as derived in section \ref{SubSec:Predicted_RM} for various tested coherence scales, $L_c$.
Column 1: The coherence scales tested;  Col. 2: $3 \sigma$ upper bound on the combined quantity  $f^{1/2} \  \overline{n_{e,fil}} \ B_{\parallel, fil, 0}$, for the associated coherence scale;  Col. 3: $3 \sigma$ upper bounds on $B_{\parallel, fil,0}$ assuming $f = 0.06$ and  $\overline{n_{e,fil}} = 10^{-5} \ \mathrm{cm^{-3}}$ for the associated coherence scales;  Col. 4: The average upper bound RM induced by filaments in foreground large-scale structure, calculated by taking the average of each individual $RM_{fil,i}$.  The uncertainty on this quantity is given by calculating the $1\sigma$ standard deviation.}
\label{Table:B_bounds}
\end{table*}

\begin{figure}
    \centering
    \includegraphics[width=0.45\textwidth]{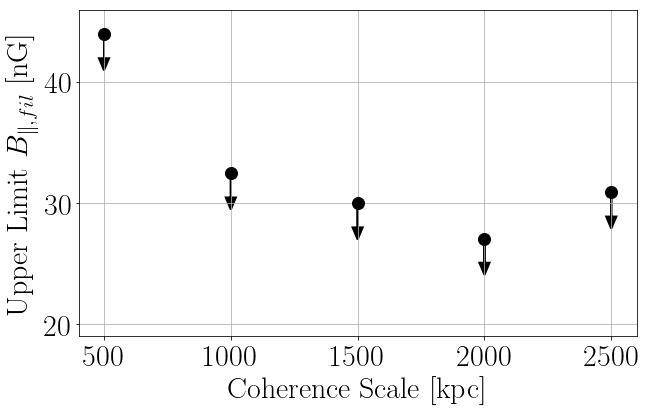}
    \caption{Upper bounds on the IGMF for all coherence scales tested in this study, the exact values can be found in Table \ref{Table:B_bounds}.}
    \label{fig:Blim_Lc_allscales}
\end{figure}

We can use equation \ref{eqn:RM_model_full} to calculate the predicted $RM_{fil}$ along the sight-lines of the 1742 observed RM sources used in this study.  As in Section \ref{SubSec:CC_Function}
 using equation \ref{ccf_eq},
 we can calculate the cross-correlation between the 1742 predicted $RM_{fil}$s and the foreground galaxy densities.  Since $B_{\parallel, fil, 0}$ is an input to equation \ref{eqn:RM_model_full}, we determine the $B_{\parallel, fil, 0}$ required to scale the resulting cross-correlation above the $3\sigma$ significance levels (as calculated in Section \ref{SubSec:Uncertainties}).  

The cross-correlation is measured at 8 linearly spaced radial bins extending to 2.5 Mpc.  Therefore we chose to scale the resulting predicted cross-correlation to $3\sigma$ at the radial bin closest in value to the input coherence scale ($L_c$) of the magnetic field. The input $B_{\parallel, fil, 0}$ to produce $RM_{fil}$s that correlate above $3 \sigma$ in the radial bin closest in value to $L_c$ is the upper bound on the IGMF. We then can determine the average RM induced by the IGMF of foreground filaments, $\left<RM_{fil}\right>$, by taking the ensemble average of each individual $RM_{fil}$ over all 1742 sight-lines.

In order to do this calculation, we must assume values for the mean electron number density in filaments ($\overline{n_{e,fil}}$), the coherence scale of the IGMF ($L_c$), and filling factor ($f$). Alternatively, from the model RM we derived in equation \ref{eqn:RM_model_full}, we can place bounds on the combination of these parameters $f^{1/2} \ \overline{n_{e,fil}} \ B_{\parallel,fil,0}$.

Because the origin of these fields are unknown and no previous direct observations exist, we chose to test a set of 5 coherence scales: 0.5, 1.0, 1.5, 2.0, and 2.5 Mpc.  We cannot probe scales larger than 2.5 Mpc as our correlation scale does not extend further than its largest radial bin. We find that the upper bound on $f^{1/2} \ \overline{n_{e,fil}} \  B_{\parallel,fil,0}$ is similar for all scales tested, all on the order of $\sim 10^{-8}\ \mathrm{ \mu G \ cm^{-3}}$. For the largest coherence scale of $2.5 \ \mathrm{Mpc}$, we obtain an upper bound of $f^{1/2} \ \overline{n_{e,fil}} \  B_{\parallel,fil,0} <  7.6 \times 10^{-8}\ \mathrm{ \mu G \ cm^{-3}}$, corresponding to an average RM enhancement of 4.7 $\mathrm{rad/m^2}$ due to filaments in the observer frame. The upper bounds for the other coherence scales are found in column 2 of Table \ref{Table:B_bounds}, these upper bounds are plotted in Figure \ref{fig:Blim_Lc_allscales} as a function of coherence scales for a visual representation.

To further constrain $B_{\parallel, fil, 0}$, we make assumptions about $f$ and $\overline{n_{e,fil}}$.  The filling factor of filaments within the IGM is not well constrained. Simulations have shown that $f=0.06$ at lower redshifts ($z<0.5$) \citep{Aragon2010,Cautun2014,Dave2010}.  
The average free electron density within filaments is also not well constrained because filaments are difficult to directly observe \citep{Cantalupo2014}; most values on the average electron densities come from simulations. These studies quote values from $10^{-4}$ to $10^{-6} \ \mathrm{cm^{-3}}$ \citep{Cen2006,Ryu2008,Akahori2010}, as in \cite{Osullivan2018} here we adopt a typical value $\overline{n_{e,fil}} \approx 10^{-5} \mathrm{cm^{-3}}$.

The upper bounds of $B_{\parallel,fil,0}$ for the various coherence scales are found in column 3 of Table \ref{Table:B_bounds}. For the largest scale we test, and the above values used for $\overline{n_{e,fil}}$ and $f$ we obtain a $3 \sigma$ upper bound on the IGMF to be $31 \ \mathrm{n G}$ for a coherence scale of $2.5 \ \mathrm{Mpc}$. We find little difference between the IGMF upper bound between 1.0 to 2.5 Mpc.  We found that the $3 \sigma$ upper bound on the IGMF for the smallest coherence scale of $0.5 \ \mathrm{Mpc}$ is $44 \ \mathrm{nG}$. This value is much less constraining due to the fact that the error bars at this scale are larger, and therefore not ideal to use as an upper bound compared to the other scales.

The associated predicted $RM_{fil}$s generated using equation  \ref{eqn:RM_model_full} can be seen in the left panel of Figure \ref{fig:predicted_RRM_CCF_distribution}, in which we also compare to the observed RRMs used to calculate the cross-correlation in Section \ref{SubSec:CC_Function}. 
The right panel of Figure \ref{fig:predicted_RRM_CCF_distribution} shows the resulting cross-correlation function calculated using the predicted $RM_{fil}$s, this output correlation is greater than $3\sigma$ at the largest radial bin - allowing us to use the associated $B_{\parallel, fil, 0}$ value as an upper bound.

\subsubsection{Effects of Model Assumptions on Upper Bounds}



To ensure that the choice of weighting functions and parameters produced robust upper limits, we tested scaling length values, $r_c$, between 0.5 Mpc and 2.5 Mpc, and found that the final upper bound $B_{\parallel,fil,0}$ results were consistent within $\pm10\%$ for $L_c = 1 - 2.5 \ \mathrm{Mpc}$, and within $\pm30\%$ for $L_c = 0.5 \ \mathrm{Mpc}$, for all $r_c$ values tested.  We also found that the choice of a \cite{King1972} profile matched the upper bounds results from using a Gaussian profile, within $\pm15\%$.  Therefore we conclude that the choice of weighting function and scaling lengths are not a source of major uncertainty in our methods. 

We assumed a cylindrical symmetry for filaments, along with mass and flux conservation, to derive equation \ref{eqn:Bpara_model}. \cite{Blasi1999} used the same assumptions as our study, although invoked spherical symmetry, to derive $B \sim n_e^{2/3}$ (a relation used in IGMF studies such as \citealt{Pshirkov2016,OSullivan2020}). When using the $B \sim n_e^{2/3}$ scaling relation in our study, we obtain upper bound results that are $\sim15\%$ lower than those quoted in Table \ref{Table:B_bounds}. The scaling relation $B \sim n^{k}$ of magnetic fields with various gas densities has been previously explored in the literature (see \citealt{Vallee1995} for a thorough review).  Using measurements from previous studies, \cite{Vallee1995} concludes that for low gas densities $n_e < 100 \ \mathrm{cm^{-3}}$, $k=0.17\pm0.03$.  When using this scaling relation in our model, we find that the upper bound results increase by $\sim40\%$.
While the exponent of the magnetic field and density scaling does have an implication on our results, it is not significant.

The upper bounds on the combined factor $f^{1/2} \overline{n_{e,fil}} B_{\parallel,fil,0}$ are more robust than those for $B_{\parallel,fil,0}$ because we include further assumptions of values of $\overline{n_{e,fil}}$ and $f$. We tested various filling factors to understand the effect on our resulting upper bounds. In our calculations, we assumed $f = 0.06$.  For a lower value $f=0.01$, and for a higher value $f=0.2$, we find an increase by $\sim 145\%$ and a decrease by $\sim 45\%$ in the upper bound on all scales, respectively.

As discussed in Section \ref{SubSec:generating_upper_bounds}, simulations have shown that $\overline{n_{e,fil}}$ can vary between $10^{-4} \ \mathrm{cm^{-3}}$ and $10^{-6} \ \mathrm{cm^{-3}}$. In our quoted upper bounds, we assumed $10^{-5} \ \mathrm{cm^{-3}}$. Changing $\overline{n_{e,fil}}$ in equation \ref{eqn:ne_model} to test the extreme range for filaments, we find that the resulting upper bounds increase and decrease by an order-of-magnitude for $\overline{n_{e,fil}} = 10^{-6} \ \mathrm{cm^{-3}}$ and $\overline{n_{e,fil}} = 10^{-4} \ \mathrm{cm^{-3}}$, respectively, for all $L_{c}$ values.  While the changes in both $f$ and $n_{e}$ cause a more drastic change in the upper bound values we obtain, these are also the least well-constrained values used within our study.

\begin{table*}

\caption{A selection of relevant observational upper bounds on large scale extragalactic magnetic fields within the filamentary IGM, sorted by increasing IGMF upper bound.}

\begin{tabular}{|l|l|l|l|l|}

\hline

\textbf{Upper Bound} & \textbf{Coherence Scale ($L_c$)} & \textbf{Redshift Range} & \textbf{Reference} & \textbf{Technique Summary} \\ \hline



0.65 nG & Jean's scale ($\lambda_j (z)$) & $0 < z < 5$ & \cite{Pshirkov2016} & RM vs. z\\ 

1.7 nG  & $1/H_0$ &  &  &  \\ \hline

1 nG & $1/H_0$ & $0 < z < 2.5$ & \cite{Blasi1999} &  RM vs. z\\ 

 6 nG & 50 Mpc&  &  &  \\ \hline

4 nG & $L_c/\lambda_j(z=0) \leq 1000$ & Unknown\footnotemark & \cite{OSullivan2020} & $\mathrm{\Delta RM}$ between adjacent sources\\ \hline

21 nG & $\sim$Gpc\footnotemark& $z \sim 0.1$ & \cite{Ravi2016} & RRM from FRB 150807\\ \hline

27 nG - 44 nG & 2.5 Mpc - 0.5 Mpc & $0.1 < z < 0.5$ & This paper & RRM cross-correlation\\ \hline

30 nG & $\theta = 1^{\circ}$\footnotemark & $z < 0.048$  & \cite{Brown2017} & Synchrotron cross-correlation\\ \hline

30 nG - 60 nG & 2 Mpc & $<z>= 0.14 \pm 0.01$  & \cite{Vernstrom2021} & Statistically Stacking Filaments\\ \hline

30 nG - 1980 nG & 1 - 4 Mpc & $0 < z < 0.57$ & \cite{Vernstrom2017} & Synchrotron cross-correlation\\ \hline

37 nG & 1 Mpc &  $0 < z < 1$ & \cite{Vernstrom2019} & $\mathrm{\Delta RM}$ between adjacent sources\\ \hline

250 nG & $\sim$10 Mpc &  $0.095 < z < 0.142$ & \cite{Locatelli2021} & Non-detection of filament accretion shocks\\ \hline

300 nG & 300 kpc & $z \sim 0.34$ & \cite{Osullivan2018} & Observed intervening filament\\ \hline

300 nG & 0.5 Mpc & $0.0304 < z < 0.0414$\footnotemark & \cite{Xu2006} & RM correlation with galaxy counts\\ \hline
\end{tabular}

\flushleft{$^5$ Redshift information not used \\
$^6$ Averaged over full sight-line \\
$^7$ Cross-correlation computed on an angular scale \\
$^8$ The redshift range of the Hercules super-cluster was determined by \cite{Kopylova2013}}
\label{tab:past_B_bounds}
\end{table*}

\subsection{Comparison To Other Studies}\label{SubSec:Other_studies_compare}

It is difficult to make direct comparisons between upper bound studies, as the techniques and data used differ greatly. Also, knowledge of many of the values we assume ($n_e$,$L_c$, and $f$) in our study are not robustly known, and thus values assumed vary between studies. Techniques to place bounds come from statistical methods, observations, and simulations. Furthermore, because of the lack of observations, simulations have been used to place bounds on the IGMF. These range from simulated observables for forecasting future radio polarimetric surveys, to full universe magneto-hydrodynamic (MHD) simulations.

\cite{Vazza2017} used the MHD \texttt{ENZO} \citep{Bryan2014} code and incorporated galaxy formation, dark matter, gas, and chemical abundances, to simulate cosmic magnetic field evolution from $z=38$ to $z=0$.  Whereas most cosmic simulations focus on a single magneto-genesis model, \cite{Vazza2017} incorporate both primordial and astrophysical seeding models to compare output observables (such as Faraday rotation, synchrotron emission, cosmic ray propagation, etc.).  They found that the magnetic fields in higher density regions such as within galaxy clusters and groups were on the order of $\mathrm{\mu G}$ for all scenarios, whereas in the less dense filament environments, the magnetic field strength strongly depends on the magneto-genesis models assumed.  For purely astrophysical phenomena, they found filaments to be magnetized at the $\mathrm{nG}$ level, whereas for purely primordial fields magnetic fields are found to be $\sim 0.1 \ \mathrm{nG}$, and for dynamo amplification models they find fields on the order of $100 \ \mathrm{nG}$.  Our upper-bound values possibly disfavour the dynamo amplification models from \cite{Vazza2017}, for which the simulations start with a weak seed field of $10^{-9} \mathrm{nG}$ and allow for dynamo amplification of these fields through solenoidal turbulence within large scale structure. However, in general the magnetic field bounds we have placed in this study are not stringent enough to be able to discern between pure primordial vs. pure astrophysical models.


In Table \ref{tab:past_B_bounds}, we list upper bounds on the IGMF from relevant observational studies.  We will discuss a subset of them in further detail below.

Observationally, upper bounds placed on the IGMF are extremely limited - this is mostly due to the estimated low electron densities in filaments \citep{Cen2006}. Notably, observing a giant double-lobed radio galaxy, \cite{Osullivan2018} determined that one of the galaxy's radio lobes passed through an IGM filament, allowing them to single out the RM due to the filament.  This RM corresponded to a density weighted upper bound of $\sim 0.3 \ \mathrm{\mu G}$.  This is consistent with, but less constraining then, the upper limit we have derived here.


\cite{Xu2006} provided IGMF upper bounds from a large statistical study using RM data derived from \cite{Simard-Normandin1980} (with and without redshift information), and LSS information from galaxy counts in the Hercules and Perseus-Pisces superclusters (from the second Center for Astrophysics survey, CfA2, and the Two Micron All Sky Survey, 2MASS, galaxy catalogues). 
They used RMs smoothed at a scale of $7^{\circ}$ (smoothed RMs, SRM) to study very large scale variations.  
\cite{Xu2006} then looked for averaged correlations across known super-cluster structures between the averaged SRMs within the region, and the galaxy-count-weighted path-length through the structure. They use a very simple model for RMs induced from a filament, with similar but less complex components compared to our model, such as; coherence scale ($L_c$), free electrons in filaments ($\overline{n_{e,fil}}$), path length through filament structures, and $B_{\parallel,fil,0}$.

\cite{Xu2006} obtained upper bounds by comparing expectations between this simple RM model and the SRMs in the Hercules cluster.  Testing values of Lc = 200 kpc - 800 kpc and $n_{e,fil,0} = 0.5 - 2 \times 10^{-5} cm^{-3}$, they obtain upper bounds between 0.4 and 0.3 $\mathrm{\mu G}$. These bounds are larger than ours for a multitude of reasons such as: fewer RM sources and galaxies used, and not subtracting off the GRM.  This sort of correlation and analysis was done looking at averaged and smoothed values over the entire structure, while our approach calculates a cross-correlation function across all RM sight-lines.

Various other statistical methods have been used to place bounds, each with their own respective strengths and weaknesses. A large source of error with these types of analyses is accurately subtracting the Galactic foreground contribution from RMs so that we only obtain extragalactic signals (see equation \ref{RM_components}).  
If sources occupy the same region of sky, the Galactic component of the RM is comparable between the sources. \cite{Vernstrom2019} used this to determine limits for the IGMF by comparing statistical RM differences between populations of physical pairs of extragalactic polarized sources (found at similar redshifts and close on the sky) versus populations of non-physical pairs (occupying different redshifts but still close together on the plane of the sky) within the \cite{Hammond2013a} RM catalogue. The statistical differences between the RMs of these populations would be due to the extra rotation induced on the non-physical pairs from the IGMF, allowing upper limits of $40 \ \mathrm{nG}$ to be placed on the parallel component of $B_{IGM}$. This value broadly agrees with our bounds at 24 nG, using a subset of the same RM catalogue though using different techniques.

Following the same novel technique as \cite{Vernstrom2019}, \cite{OSullivan2020} used RMs of 201 physical and 148 random pairs derived from the LOFAR Two-Metre Sky Survey (LoTSS) DR1 \citep{Shimwell2019}.  They found an upper bounds on the magnetic field of the IGM of 4nG at Mpc scales. Because LoTTS is observed at 144 MHz, any RMs derived from these bands are extremely sensitive to any Faraday depolarization, therefore the authors conclude that these sources  preferentially occupy less dense and weakly magnetized regions within the IGM - such as on the boundaries of filaments and voids.  In contrast, the 1.4 GHz sample used in both \cite{Vernstrom2019} and our study are less susceptible to Faraday depolarization and the bounds on the IGM derived from these data should be more representative of denser filaments.  Because the bounds we derived use the same 1.4 GHz RM data \citep{Hammond2013a} and obtain similar magnetic field upper limit as \cite{Vernstrom2019}, we cannot disfavour the hypothesis that LoTSS sources occupy rarefied environments.


The hypothesis that LoTSS sources occupy extremely rarefied environments was further confirmed by \cite{Stuardi2020},
who looked at 37 polarized LoTTS giant radio galaxies (GRGs) - galaxies whose outer lobes extend well into the IGM and have a limiting size of 700 kpc.  Using the same approaches as \cite{Vernstrom2019} and \cite{OSullivan2020}, they analyzed the RM differences between the lobes of each of the galaxies.  They found that the RM difference between the lobes of the galaxies were consistent with variations in the local Milky Way Galactic foreground magnetic field, as was found by \cite{Vernstrom2019}.  
\cite{Stuardi2020} also found that there were small amounts of Faraday depolarization between 144 MHz and 1.4GHz,  $D_{144 MHz}^{1.4 GHz} > 0.03$ for all sources.  This can be explained by the presence of a low density medium ($n_e \sim 10^{-5} \mathrm{cm^{-3}}$) with magnetic fields on the order of $0.1 \ \mathrm{\mu G}$, tangled on kpc scales. While it is unclear whether these sources probe the same IGM component as our studies, this study showed the capabilities of using GRGs to study intergalactic magnetic fields as more data becomes available.


On scales smaller than we probe here, \cite{Lan2020} used similar cross-correlation and analysis techniques to ours, between ~$1100$ background RRMs ($z>1$, using redshift information from \citealt{Hammond2013a}, RM information from \citealt{Farnes2014}, and foreground GRM subtraction from \citealt{Oppermann2015}) and foreground galaxy distributions as sampled by the DESI Legacy Imaging Surveys \citep{Dey2019}, to extract the characteristic RM and $<B_{\parallel}>$ of the circum-galactic medium (CGM) around galaxies.  They obtained a null correlation, but used the trend between $\sigma_{RM}$ and $N_{gal}$ (number of foreground galaxies within the RRM line-of-sight) to model the random field reversals due to intersecting magnetic fields of the CGM.
They use the $3\sigma$ significance levels on $\sigma_{RRM}$ vs. $N_{gal}$ and a random walk model to place upper bounds.  They found $3\sigma$ upper bounds of $RM_{CGM} \sim 15 \ \mathrm{rad \ m^{-2}}$ and $B_{\parallel,CGM} < 2\mu G$ within the CGM virial radius (impact parameters less than 200 kpc).

Using a combination of the cross-correlation methods from \cite{Lan2020} and our study, with foreground galaxy density maps and denser RM grids, we can hope to accurately probe the magnetic fields of the CGM, and the transition region between the CGM and the IGM (the CGM is thought to extend between 100 kpc - 200 kpc from individual galaxies, \citealt{Shull2014}).  The finer precision of denser RM grids will allow to accurately measure correlations at these scales \citep{Bernet2010,Bernet2013,Farnes2014,Prochaska2019}.  Understanding the magnetic properties of this transition region is key to understanding the interplay between the properties of the IGM, CGM, and galaxies themselves \citep{Borthakur2015}.  Additionally, this region mediates the acquiring of neutral gas from the IGM to individual galaxies to further fuel star formation and other galactic processes. Furthermore, exploring the gas and metal properties of the CGM and IGM may allow us to differentiate between various galaxy formation models \citep{Tumlinson2017,Codoreanu2018}.


Cross-correlation techniques can also be used to detect the synchrotron cosmic web (which is also much too faint to directly observe), and hence constrain the perpendicular component of the IGMF in filaments. This was recently done by \cite{Brown2017} by cross-correlating S-PASS continuum data at 2.3 GHz and MHD simulations of the cosmic web (using models by \citealt{Dolag2005,Donnert2009}), placing bounds on the order of $0.1 \ \mathrm{\mu G}$ on the IGMF.  The upper bound places by \cite{Brown2017} agreed with other synchrotron based bounds placed, such as in \cite{Vernstrom2017}, and is consistent with bounds placed in our study. 

Most recently \cite{Locatelli2021} used the non-detection of synchrotron accretion shocks between two adjacent galaxy clusters located using LOFAR (pair 1: RXCJ1155.3+2324 /RXCJ1156.9+2415, and pair 2: RXCJ1659.7+3236/RXCJ1702.7+3403) by comparing their observations to MHD \texttt{ENZO} \citep{Bryan2014} simulations of galaxy clusters of similar properties.  By doing this they were able to place bounds on the intergalactic magnetic field present to $B_{fil} < 0.25 \ \mathrm{\mu G}$ for Mpc scales.  Although these bounds are consistent but less constraining than those places in our study, \cite{Locatelli2021} used only two clusters.  With more cluster observations with radio telescopes such as LOFAR this technique can prove to be powerful.

Forecasts of future radio surveys have lead to predictions for what such an IGMF signal would look like, if detected. \cite{Stasyszyn2010} found, using cross-correlations of simulated data (a similar technique to our study), that direct detections of the large-scale magnetic field that permeates the IGM in filaments will only be possible with larger RM catalogues from future telescopes. \cite{Akahori2014a} explored the effectiveness of using simulated RM grids with varying sample sizes and statistical techniques to detect magnetic fields structure in filaments.  They found that basic properties of large-scale magnetic fields can be determined statistically using upcoming surveys, allowing us to distinguish between magneto-genesis models.
For our study, it is not possible to analyze the shape of our cross-correlation function to obtain information about the coherence scale of the IGMF, nor to constrain the generation mechanism of the field, as we did not detect a significant cross-correlation signal.  Such studies must wait for future surveys.

\subsection{Sources of Uncertainty}\label{SubSec:Sources_of_Uncertainties}

\begin{figure}
	\centering
    \includegraphics[width=0.5\textwidth]{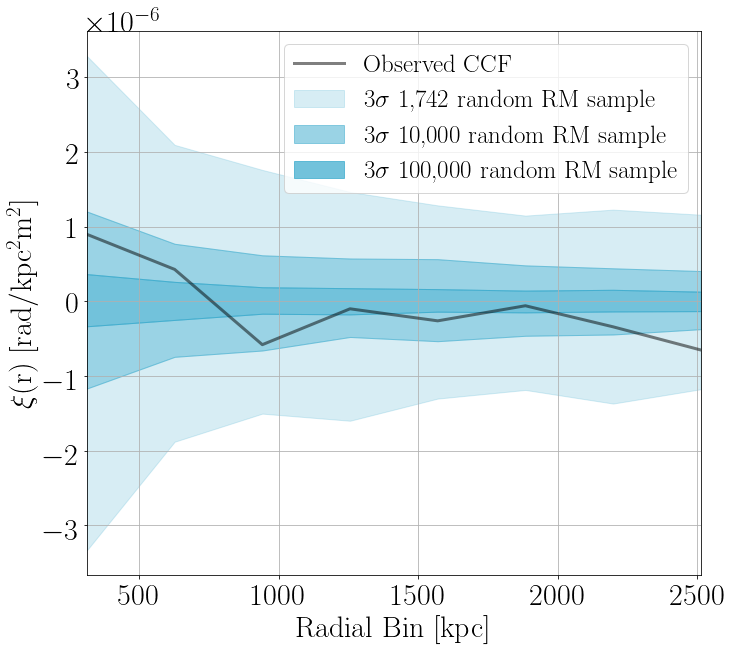}
    \caption{The $3\sigma$ significance levels of random CCFs for increasing sample sizes of RRM sources.  The RRMs were generated by generating randomized sightlines and using RRMs from real data.}  
    \label{fig:biases_sampleRRMs}
\end{figure}

Our study is mostly limited by the availability of extra-galactic RM sources with redshift information. As we progress towards an era of large data availability, such as the upcoming data releases from the Australian Square Kilometre Array Pathfinder (ASKAP) within the next few years, the bounds we can place on $B_{\parallel,fil}$ will become more robust. Once completed, ASKAP's Polarization Sky Survey of the Universe's Magnetism (POSSUM)\footnote{http://askap.org/possum} \citep{Gaensler2010} will comprise the largest catalogue of extra-galactic rotation measures to date, with RMs for one million extra-galactic radio sources with a source-density coverage of $\geq \ 25$ RMs $\mathrm{deg^{-2}}$.

To demonstrate that using this technique with a larger sample of RMs will obtain more stringent upper bounds, we generated simulated RM data using the technique described in Section \ref{SubSec:Uncertainties} to obtain sample catalogues of 1742 RMs (the size of the \citealt{Hammond2013a} sample used in this study, for comparison), 10,000 RMs, and 100,000 RMs.  We then obtain $3\sigma$ significance levels for 1000 realizations of sample catalogues of these sizes, plotted in Figure \ref{fig:biases_sampleRRMs}.  For the sample sizes larger than $1742$, we draw RM values (with replacement) from the full \cite{Taylor2009} catalogue, applying the same sample cuts as in section \ref{SubSec:Sample_Selection}, and analysis in section \ref{SubSec:residual_RM} to obtain RRMs. The error bars become smaller with increasing sample sizes.  At the largest radial bin, 2.5 Mpc, this corresponds to a $\sim$3 times better sensitivity for a sample size of 10,000 RMs than for a sample size of 1742 RMs, whereas a sample size of 100,000 RMs correspond to $\sim$9 times better sensitivity at scales of 2.5 Mpc. With the data to become available from surveys such as POSSUM, we expect that this method will far better constrain $B_{\parallel,fil}$.

\begin{figure}
	\centering
    \includegraphics[width=0.5\textwidth]{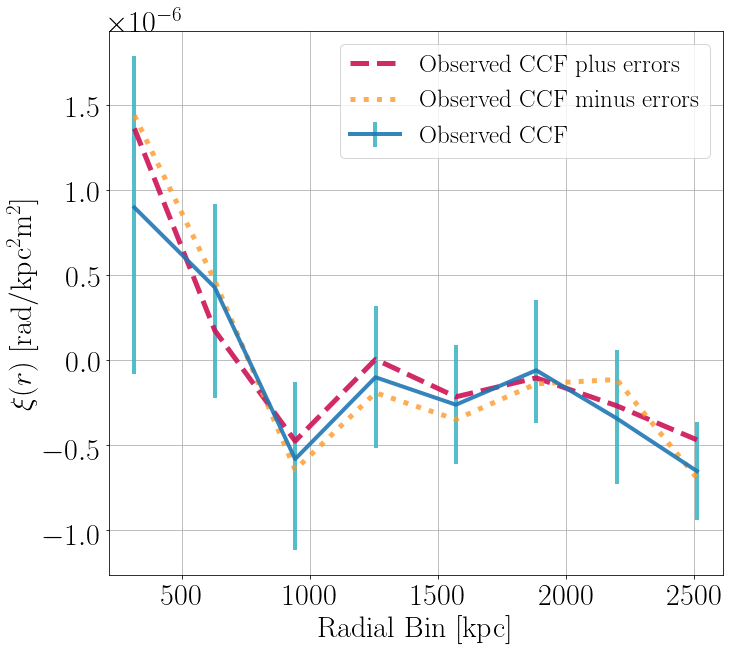}
    \caption{The resulting cross-correlation functions when we include the errors associated with the RM sources.  The red dashed line is calculated by adding a random Gaussian drawn fraction of the error to the $|RRM|$ value, while the orange line is calculated in the same manner but the errors are subtracted from the $|RRM|$ values.} 
    \label{fig:error_RRMs}
\end{figure}

The main source of uncertainties in this analysis originates from the data-sets used.  
\cite{Hammond2013a} use RMs determined from NVSS, which is determined by linearly fitting the polarization angle vs. $\lambda^2$ for only 2 narrowly spaced bands (see Section \ref{SubSec:RM_cat}), which may lead to spurious RM values for sources.

These spurious RMs could manifest in $n\pi$ ambiguities.  In this case the phase angles wrap around at $\pi$ and thus the observed polarization angle is only known to modulo $n \pi$ radians \citep{Heald2009}. Sources affected by this could have erroneously high RMs offset from the true RM by $\pm 652.9 \ \mathrm{rad/m^2}$ \citep{Ma2019a}.  \cite{Ma2019a} explored this effect in the \cite{Taylor2009} catalogue by using follow up large-bandwidth observations of suspicious sources.  They concluded that the $n\pi$ ambiguity would affect $>50$ sources in the full $\sim$37,000 sample catalogue.

Additionally, the \cite{Taylor2009} catalogue does not adequately treat Faraday thick sources.  Faraday thick sources are polarized sources whose fractional polarization varies from the linear relationship with $\lambda^2$ assumed by \cite{Taylor2009}.  This Faraday thick behaviour can be caused by many situations; regions simultaneously emitting and rotation, or various regions rotating at different amounts along the line-of-sight (see a detailed discussion in \citealt{Sokoloff1998}). Because the two bands used to produce the RMs of these sources are spaced close together, the RM information captured could be from a particular Faraday rotating structure and not representative over larger bandwidths.

However, for both of these cases without re-observing all sources at larger bandwidths, it is impossible to determine which sources would be affected. POSSUM will observe at 800 MHz to 1088 MHz at 1 MHz resolution, corresponding to $\Delta \nu$ = 288 MHz, leading to much more robust RM value assignments.
Additionally, POSSUM will use techniques such as QU-fitting \citep{OSullivan2012} and RM synthesis \citep{Brentjens2005} to extract Faraday depth information and RMs from sources. These techniques are much more reliable than angle fitting techniques used for older data-sets. 

We did not use the error bars associated with the \cite{Hammond2013a} RM sources in our cross-correlation.  To ensure that the RM errors would not have an effect on the resulting cross-correlation function, we recalculated the cross-correlation for values of RM adjusted by a Gaussian drawn value multiplied by its error.  We found that running this for 100 cases did not change the outcome of the cross-correlation function within $1\sigma$. This effect can be seen for one run in Figure \ref{fig:error_RRMs}.

\section{Conclusion}
Because of the low gas densities in cosmic web filaments, cosmic magnetic fields are difficult to detect, with estimated values in the nG regime. These weak magnetic fields require statistical approaches to extract their signal from large data-sets.

We use a cross-correlation statistical approach between 1742 background RMs (from \citealt{Hammond2013a}, $z>0.5$) and the superCOSMOSxWISE all-sky photometric redshift catalogue of 18 million galaxy sources \citep{Bilicki2016} to trace large scale structure between $0.1 < z <0.5$).  We used the \cite{Oppermann2015} Galactic rotation measure (RM) map to determine the residual rotation measures (RRMs). We then perform a cross-correlation between the background RRMs and the foreground galaxy density maps over 3 redshift bins ($0.1 <z <0.2$, $0.2 < z< 0.3$, and $0.3 <z < 0.5$) at impact parameters between $\sim 0.5$ Mpc to 2.5 Mpc.

Our results are as follows:

\begin{enumerate}[i]
  \item We found no statistically significant correlation between RRMs and the foreground galaxy density.
  \item We used a simple model and the $3\sigma$ significance levels from the observed cross-correlation to place an upper bound on the mean co-moving magnetic fields within filaments.  The $3\sigma$ upper bound for magnetic field coherence scales between 1 - 2.5 Mpc was $\sim$30 nG (the magnitude was similar for all 4 scales probed), and 44 nG for 0.5 Mpc.  These upper bounds depend on assumptions for the filling factor and mean electron density.
  \item Comparing our result to simulations of \cite{Vazza2017} tends to disfavour dynamo amplification models of magneto-genesis, though we cannot discern between primordial and astrophysical magneto-genesis models.   We found that this upper bound agreed well with upper bounds derived from other past statistical studies. A statistical study by \cite{OSullivan2020} using 144 MHz LoTSS RM data suggested a small upper bound of $<0.5$ nG, but focuses on less dense regions of the IGM and thus were placing bounds on magnetic fields in less dense media. 
\end{enumerate}

The main limitations with our method is the relatively small sample size of RMs, and the potentially unreliable way the RMs were determined in the NVSS-derived catalogue.  Future data-sets in the ASKAP and SKA era will allow for more robust bounds to be placed on these elusive intergalactic magnetic fields.

\section*{Acknowledgements}
We thank the anonymous referee for their useful comments. We also thank Lawrence Rudnick, Franco Vazza, Dongsu Ryu, Yik Ki (Jackie) Ma, Shane O'Sullivan, Ue-Li Pen, Cameron Van Eck, Mubdi Rahman, Norm Murray, and Matthew Young for useful discussions. 

This research has made use of the SIMBAD database,
operated at CDS, Strasbourg, France, and the NASA/IPAC Extragalactic Database (NED), which is funded by the National Aeronautics and Space Administration and operated by the California Institute of Technology.

The Dunlap Institute is funded through an endowment established by the David Dunlap family and the University of Toronto. We acknowledge the support of the Natural Sciences and Engineering Research Council of Canada (NSERC) through grant RGPIN-2015-05948, and of the Canada Research Chairs program.

This work was performed traditional land of the Huron-Wendat, the Seneca, and most recently, the Mississaugas of the Credit River. 

\section*{Data Availability}\label{Sec:Data_availability}

The \cite{Hammond2013a} RM-redshift data underlying this article was accessed from  \url{https://web.archive.org/web/20170519211023/http://www.sifa.org.au/Main/RMCatalogue}, the \cite{Oppermann2015} Galactic RM data was accessed from  \url{https://wwwmpa.mpa-garching.mpg.de/ift/faraday/}, and the \cite{Bilicki2016} galaxy catalogue was accessed from the Wide Field Astronomy Unit at the Institute for Astronomy, Edinburgh at \url{http://ssa.roe.ac.uk/WISExSCOS}. 
The derived data generated in this research will be shared on reasonable request to the corresponding author.




\bibliographystyle{mnras}

\bibliography{cosmic_mag} 







\bsp	
\label{lastpage}
\end{document}